\documentclass[useAMS,usenatbib]{mn2e}

\usepackage{color}
\usepackage{lscape,graphicx}
\usepackage{amsmath}
\usepackage{journal_shortcuts}
\usepackage{amssymb}

\def\cm{{\rm\thinspace cm}}

\def\erg{{\rm\thinspace erg}}

\def\ga{{\rm\thinspace gauss}}

\def\keV{{\rm\thinspace keV}}
\def\km{{\rm\thinspace km}}

\def\Mpc{{\rm\thinspace Mpc}}
\def\Msun{\hbox{$\rm\thinspace M_{\odot}$}}

\def\s{{\rm\thinspace s}}
\def\ps{{\rm\thinspace s^{-1}}}

\def\keVscm{\hbox{$\keV\cm^{2}\,$}}

\def\ergpscmps{\hbox{$\erg\cm^{-2}\s^{-1}\,$}}

\def\ergps{\hbox{$\erg\s^{-1}\,$}}

\def\kmps{\hbox{$\km\ps\,$}}
\def\kmpspMpc{\hbox{$\kmps\Mpc^{-1}\,$}}

\def\pscm{\hbox{$\cm^{-2}\,$}}

\def\pccm{\hbox{$\cm^{-3}\,$}}
\def\pscm{\hbox{$\cm^{-2}\,$}}

\def\j1431{\hbox{MaxBCG J217.95869+13.53470}}

\def\hi{\hbox{{\rm H{\sc i}}}}

\def\h0{\hbox{{\rm H}$^0$}}
\DeclareMathAlphabet{\vib}{OML}{cmm}{m}{it}

\title[A merger-revived radio phoenix in \j1431]{Evidence for a merger-revived radio phoenix in \j1431}
\author[G.~A.~Ogrean et al.]{G.~A.~Ogrean$^{1}$\thanks{E-mail:
g.ogrean@jacobs-university.de}, M.~Br\"{u}ggen$^{1}$, R. van Weeren$^{2}$, A. Simionescu$^{3}$, H.~R\"{o}ttgering$^{2}$, \and J.~H.~Croston$^{4}$\\
$^{1}$Jacobs University Bremen, Campus Ring 1, Bremen 28759, Germany\\
$^{2}$Leiden Observatory, Leiden University, P.O. Box 9513, NL-2300 RA Leiden, Netherlands\\
$^{3}$KIPAC, Stanford University, 452 Lomita Mall, Stanford, CA 94305, USA\\
$^{4}$School of Physics and Astronomy, University of Southampton, Southampton, SO17 1SJ}

\begin{document}

\date{Accepted 2011 January 31.  Received 2011 January 27; in original form 2010 December 23}

\pagerange{\pageref{firstpage}--\pageref{lastpage}} \pubyear{2002}

\maketitle

\label{firstpage}

\begin{abstract}

We use \emph{XMM-Newton} observations of the galaxy cluster \j1431 to analyze its physical properties and dynamical state. \j1431 is found at a redshift of 0.16, has a mass of $\sim 1\times 10^{14}$ \Msun, and a luminosity of $7.9\times 10^{43}$ \ergps. The temperature map shows the presence of hot regions towards the north and west of the brightest cluster galaxy (BCG). From the entropy distribution, regions of high entropy match the location of the hot regions; more high entropy regions are found to the west, and $\sim 165$ kpc to the southwest of the central AGN. A second X-ray bright galaxy is visible $\sim 90$ kpc to the northeast of the BCG, at a redshift of $0.162$. This galaxy is likely to be the BCG of a smaller, infalling galaxy cluster. The mass of the smaller cluster is $\sim 10$\% the mass of \j1431, yielding an impact parameter of $\sim 30-100$ kpc. We compare the results of our X-ray observations with \emph{GMRT} observations of the radio source VLSS J1431.8+1331, located at the center of the cluster. Two sources are visible in the radio: a central elongated source that bends at its northern and southern ends, and a southwestern source that coincides with a region of high entropy. The radio sources are connected by a bridge of faint radio emission. We speculate that the southwestern radio source is a radio relic produced by compression of old radio plasma by a merger shock.

\end{abstract}


\section{Introduction}

Studies of large-scale structure (LSS) formation have found that nearly all massive clusters have undergone several mergers in their history, and that present clusters are still in the process of accreting matter. A significant fraction of the accreting matter is in the form of smaller clusters and galaxy groups.

Observations have shown that $10-35\%$ of massive clusters host diffuse radio sources \citep{Giovannini1999,Venturi2007,Venturi2008}. In general, these diffuse radio sources are only found in clusters that display significant substructure in the galaxy distribution and in the thermal X-ray emitting gas \citep[e.g., ][]{Schuecker2001,Venturi2008}. These clusters are also characterized by high X-ray luminosities and temperatures. All of these signatures are indicative of the cluster undergoing a merger.

The diffuse radio sources observed are also referred to as radio halos or relics. Radio halos result from the reacceleration of a relatively old electron population as a result of galaxy cluster mergers. Radio relics are irregularly shaped radio sources with sizes ranging from 50 kpc to 2 Mpc. They can be divided into two groups \citep{Kempner2004}. Radio gischt are large elongated, often Mpc-sized radio sources located in the periphery of merging clusters. They probably trace shock fronts in which particles are accelerated via diffusive shock acceleration 
\citep[DSA; e.g., ][]{Bell1978a,Bell1978b,Blandford1978,Drury1983,Blandford1987,Jones1991}. According to DSA, if the particle acceleration is still ongoing, the integrated radio spectrum should follow a single power-law. The second category of relics are radio phoenices. In these objects, fossil radio plasma from a former activity cycle of a FRI/II radio galaxy is compressed by a merger shock wave \citep{Ensslin2001,Ensslin2002}. When a fossil radio cocoon is passed by a cluster merger shock wave, the cocoon is compressed adiabatically and not shocked because of the much higher sound speed within it. Therefore, shock acceleration cannot be the mechanism that re-energizes the relativistic electron population. However, the energy gained during the adiabatic compression combined with the increase in the magnetic field strength can boost the radio emission from the fossil radio cocoon. One prerequisite for this is that the electron population is not older than $0.2-2$ Gyr, depending on the ambient pressure of the cocoon \citep{Ensslin2001}. The radio plasma in such a radio phoenix has a steeply curved spectrum due to synchrotron and inverse Compton (IC) losses. Proposed examples of these are those found by \cite{Slee2001} and \citep{vanWeeren2011}.

The radio source VLSS J1431.8+1331, located in the galaxy cluster MaxBCG J217.95869+13.53470, at a redshift of 0.16, is a good candidate for a shocked radio plasma due to its extreme spectral index and disrupted morphology. VLSS~J1431.8+1331 was included in a sample of diffuse steep spectrum sources selected from the 74~MHz VLSS survey \citep{vanWeeren2009}. The spectral index of the source is very steep with $\alpha=-2.03 \pm 0.05$ between 74 and 1400~MHz. GMRT observations showed the source to consist of two main components (see Fig. \ref{fig:clusterimage}). The brightest eastern component is associated with the BCG and can be classified as a radio AGN. For the western component no obvious optical counterpart is detected. The western component is located at a distance of about 165~kpc from the cluster centre and has a largest linear size (LLS) of 125~kpc. The radio plasma from the AGN appears to be affected by the ICM because it bends eastwards at the northern and southern ends of the source. GMRT 325~MHz observations \citep{vanWeeren2011} reveal a faint bridge of radio emission connecting the two components suggesting a link between them. The spectral index of the radio core of the AGN is $-0.5$. The spectral index steepens to $\alpha < -2$ at the northern and southern ends of the radio AGN. The western component has a very curved radio spectrum with $\alpha=-1.5$ between 325 and 610~MHz, steepening to $\alpha=-2.5$ between 610 and 1425~MHz. The very steep curved radio spectrum indicates that the radio plasma has undergone significant synchrotron and inverse Compton losses. It is likely that this radio plasma originated from the central AGN given the faint radio bridge connecting the two components. 

Here we report results of \emph{XMM-Newton} observations of MaxBCG J217.95869+13.53470, obtained with the European Photon Imaging Cameras (EPIC). The spectra allow for a detailed spatial analysis of the distribution of physical properties (e.g. temperature, electron number density, pressure, entropy) within the central region of MaxBCG J217.95869+13.53470. Comparing the X-ray results with the results obtained from GMRT observations of the radio source allows us to study the interaction between the BCG, the radio plasma and the ICM, as well as the cluster's merging history.

Section \ref{sec:data_reduction} describes the observations and the X-ray data reduction process, while Section \ref{sec:results} presents the results of the spectral analysis. Section \ref{sec:discussion} provides an interpretation of the X-ray and radio results. Finally, Section \ref{sec:conclusions} summarizes the paper.

Unless stated otherwise, we use a flat $\rm{\Lambda CDM}$ universe with $\Omega_M=0.3$, $\Omega_\Lambda = 0.7$, and $H_0 = 70$ \kmpspMpc. At a redshift of 0.16, 1 arcsec corresponds to 2.758 kpc.

\section{Data reduction and spectral analysis}
\label{sec:data_reduction}

\subsection{X-ray data reduction}

On 2009 May 25, MaxBCG J217.95869+13.53470 was observed for $~45$ kiloseconds (ks) with the EPIC MOS and pn cameras of \emph{XMM-Newton}. The detectors were operated in full-frame mode, using the medium filter. 

The data was reduced using the \emph{XMM-Newton} Science Analysis System (SAS), version 9.0.0\footnote{http://heasarc.gsfc.nasa.gov/docs/xmm/abc/}. For each of the three detectors, we extracted two lightcurves: one in the hard band, $10-12$ keV ($12-14$ keV for pn), and one in the soft band, $0.3-10$ keV. These light curves were binned to 100 and 10 seconds, respectively. Time periods when the count rate deviated from the mean by more than $3\sigma$ were excluded from the data. The higher bin size used for binning the hard band light curves is due to the lower count rate in this energy band. Nevertheless, excluding periods with high count rates in the hard band eliminates most of the flaring. Repeating the procedure in the $0.3-10$ keV energy band further elimantes any short flares that might have been previously missed. The resulting net effective exposure times were $~41$ ks for MOS1, $41.3$ ks for MOS2, and $~35.7$ ks for pn.

Background subtraction was performed using blank sky observations developed at Leicester University by the EPIC Blank Sky team \cite[based on the work of ][]{Carter2007}, and filter-wheel closed (FWC)~\footnote{http://xmm2.esac.esa.int/external/xmm\_sw\_cal/background/\\filter\_closed/index.shtml} datasets, which were filtered in the same way as the target datasets. A composite background for each detector was created from the blank sky and FWC datasets by normalizing the FWC event rates to the difference between the target and blank sky event rates outside the field of view in the energy range $10-12$ keV (MOS) and $12-14$ keV (pn). This compensates for differences in the particle background between the target and background observations.

As customary, out-of-time events were scaled by a factor of $6.3\%$\footnote{http://www.mpe.mpg.de/xray/wave/xmm/cookbook/\\EPIC\_PN/ootevents.php} and subtracted from the pn data.

\begin{figure}
 \center
 \includegraphics[scale=0.4,angle=90]{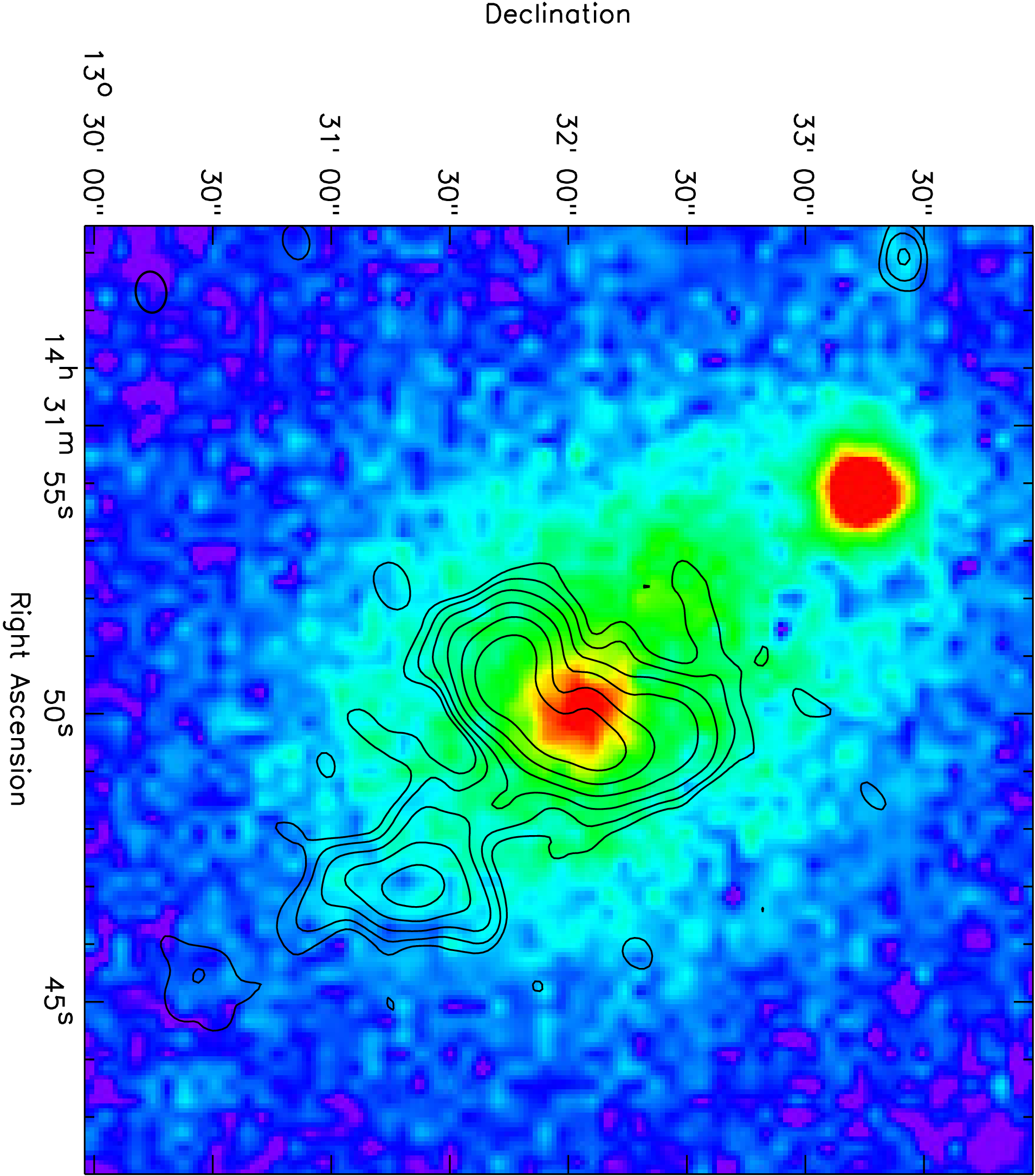}
 \caption{Surface brightness map of MaxBCG J217.95869+13.53470 in the $0.4-7$ keV energy range. The image is corrected for vignetting and the background was subtracted. GMRT 325 MHz radio contours are drawn at 4$\times$rms$\times [1,2,4,8,...]$, with ${\rm rms} = 1.3212\times 10^{-4}$ Jy\,beam$^{-1}$. \label{fig:clusterimage}}
\end{figure}

A vignetting-corrected and background-subtracted flux map of the cluster, created by combining the images in the $0.4-7$ keV band from all three detectors, is shown in Fig.~\ref{fig:clusterimage}, overlaid with GMRT 325 MHz radio contours. The X-ray emission is elongated in the NE-SW direction, parallel to the radio emission. The BCG appears slightly offset from the radio AGN. The bright source in the northeast is identified in the SDSS DR7 as SDSS J143153.78+133315.1, an AGN at a redshift of 0.16, the same as the cluster. This AGN was excluded from the spectral analysis.

\subsection{Temperature map}
\label{ssec:tmap}

To obtain a more detailed image of the physical properties of the cluster, we binned the data using an adaptive binning method based on weighted Voronoi tessellations \citep{DiehlStatler2006}, which is a generalization of the algorithm of \citet{CappellariCopin2003}. This allows us to create maps of the temperature, pressure, and entropy distributions, which are presented in the following subsections.

We binned the combined counts image to a target signal-to-noise per bin of 31 (following background subtraction) in the energy range $0.4-7$ keV, which corresponds to $\sim 960$ counts per spatial bin. Target and background spectra,  redistribution matrices (RMF) and ancillary response files (ARF) were created for each resulting bin. The target spectra were grouped to a minimum of 30 counts per spectral bin.

Using XSPEC, we fitted the spectra with a Galactically-absorbed single-temperature \texttt{APEC} model, which descibes emission from collisionally-ionized diffuse gas. The hydrogen column density, $n_{\rm H}$, was fixed to $1.52\times 10^{20}$ \pscm -- the weighted average hydrogen column density at the position of the BCG in the Leiden/Argentine/Bonn (LAB) Survey of Galactic \hi. The temperatures and metallicities were left free in the fit, but they were coupled to be the same for all detectors. The normalizations were allowed to vary independently. The temperature uncertainties in the fitted bins are between 8\% and 20\% (at the 1-sigma level), while the reduced $\chi^2$ lie between 0.5 and 1.6 (typically higher at larger radii, where the emission from larger spatial bins is not well-described by a single-temperature model).

\begin{figure*}
 \center
 \includegraphics[scale=0.29]{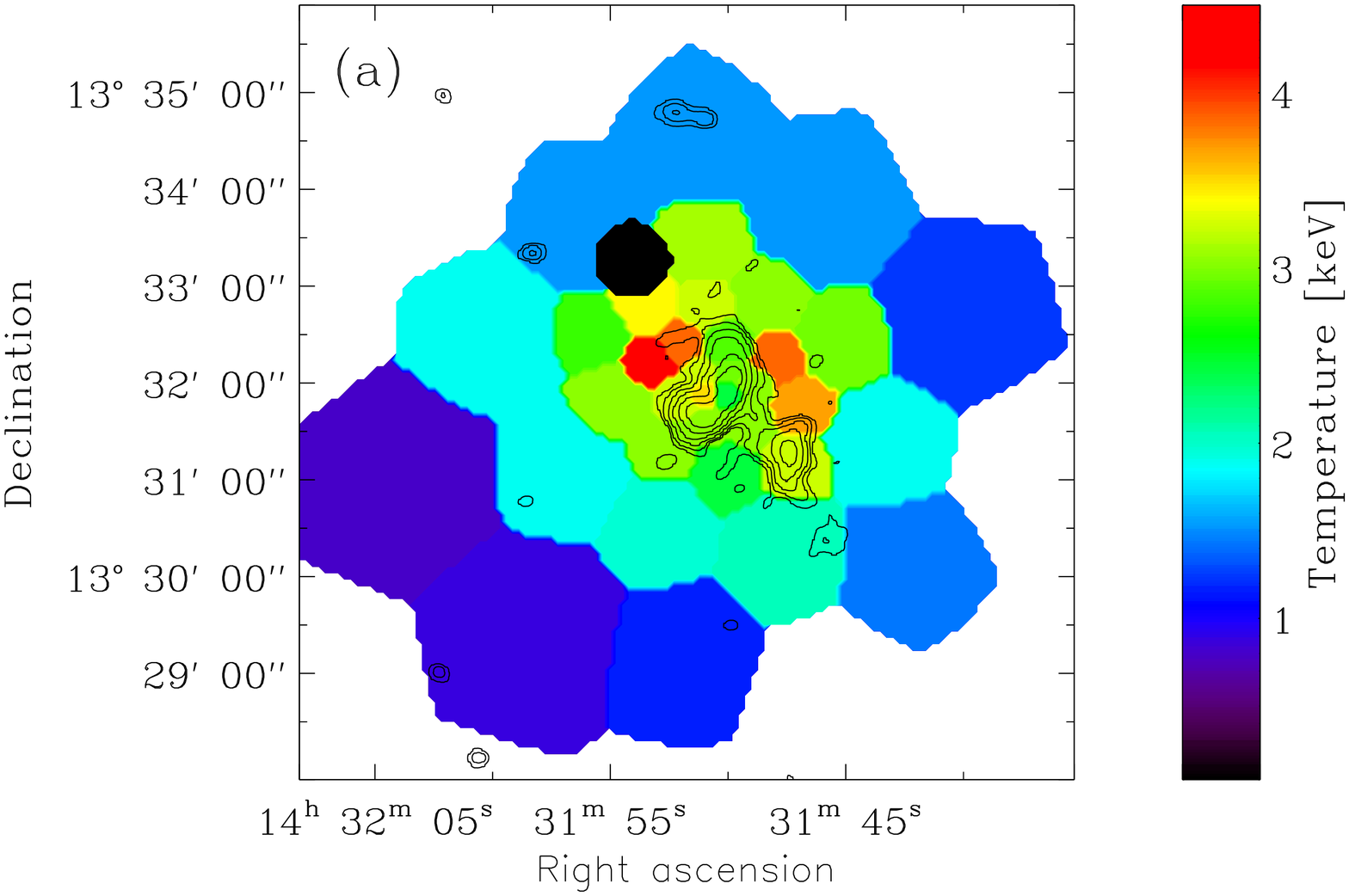}
 \includegraphics[scale=0.29]{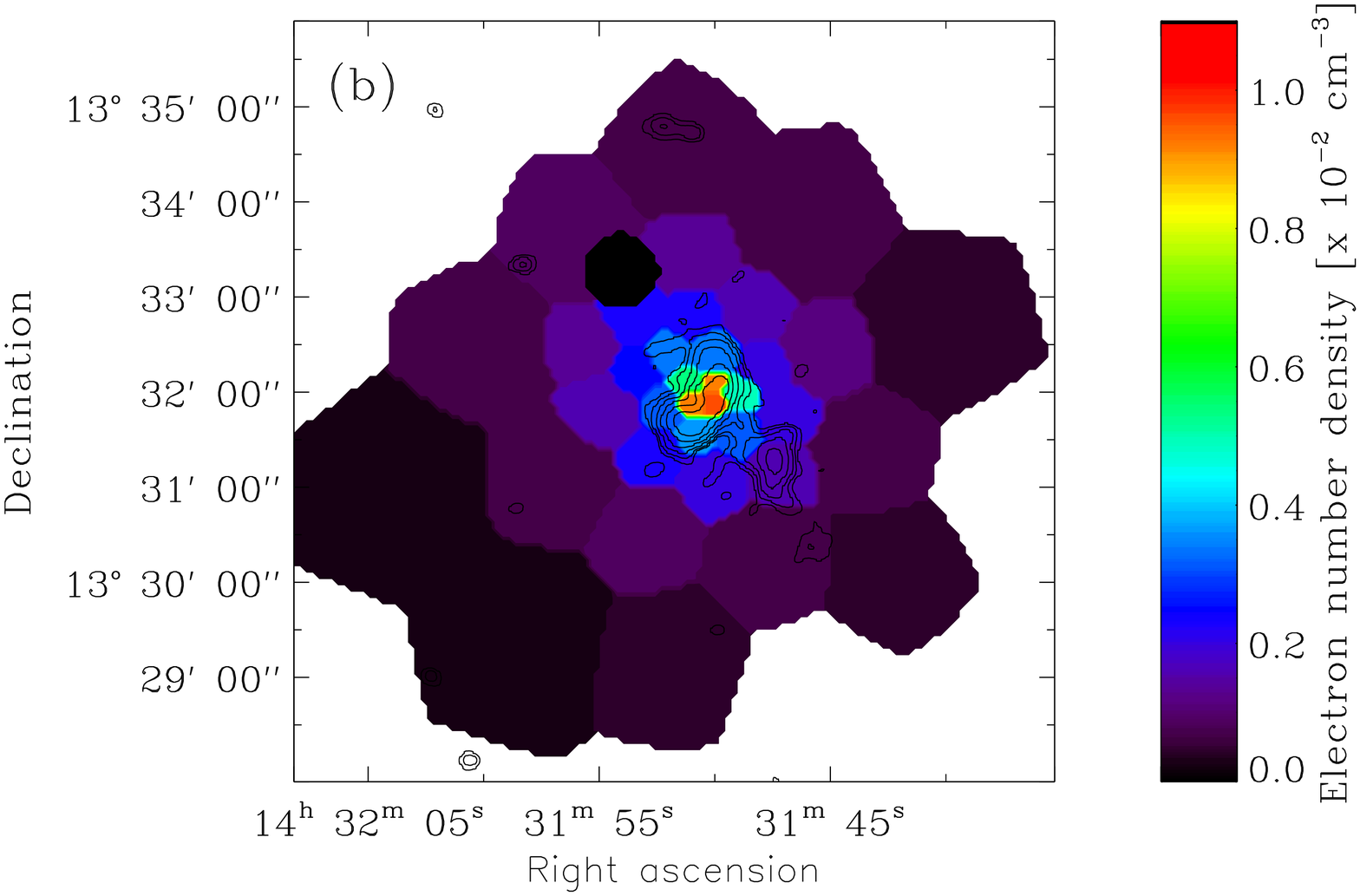}
 \includegraphics[scale=0.29]{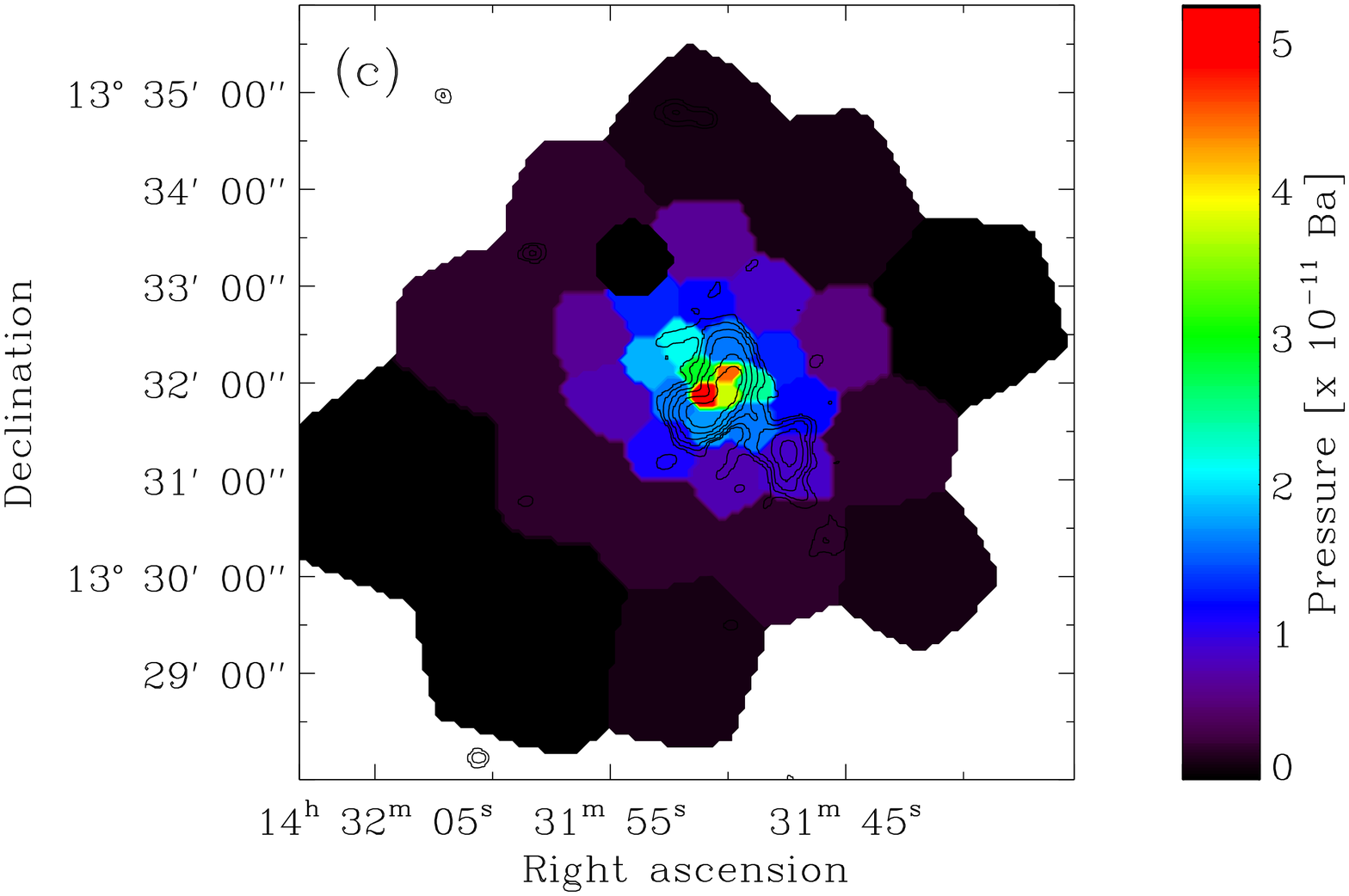}
 \includegraphics[scale=0.29]{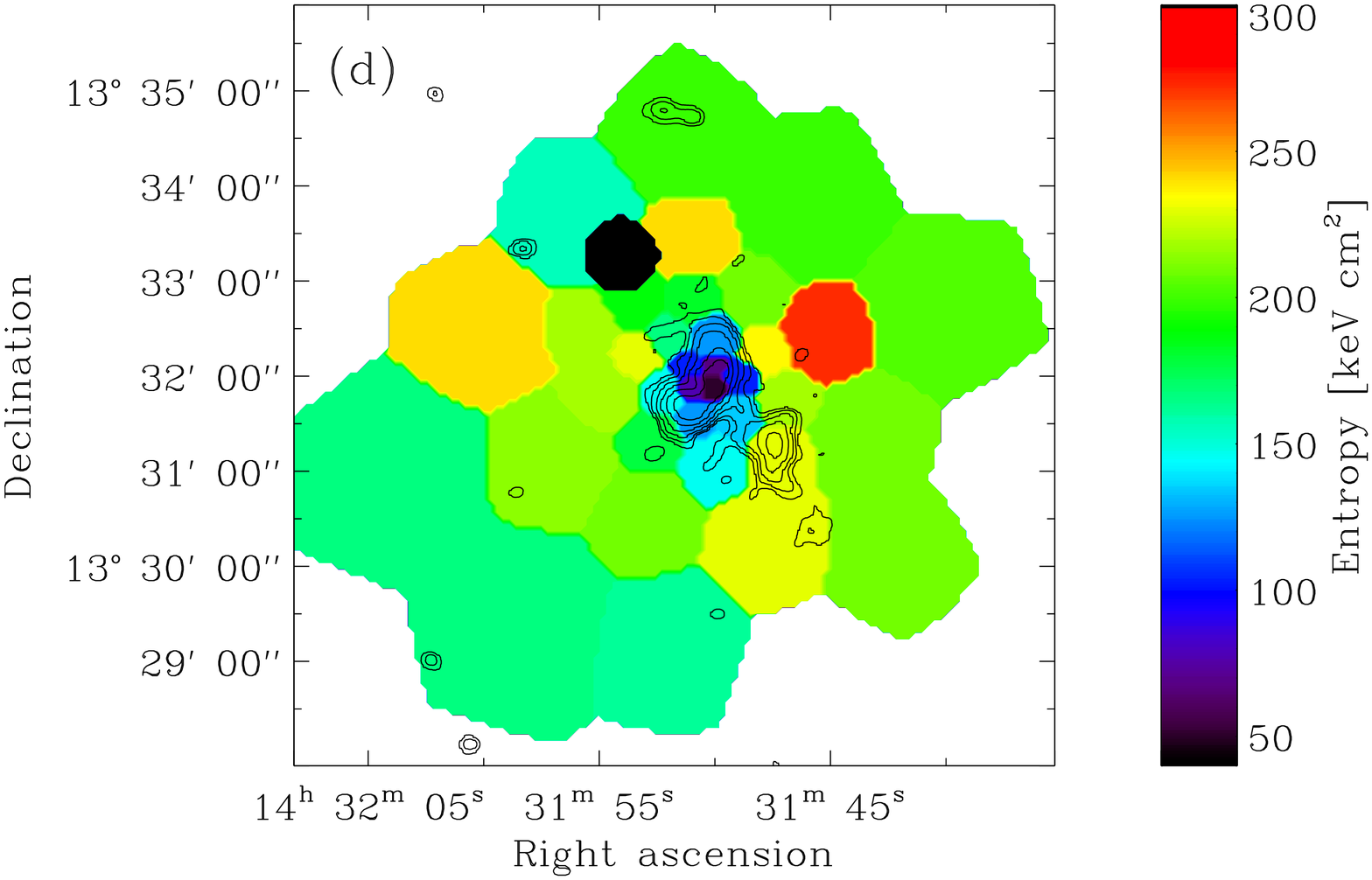}
 \caption{(a) Temperature map of MaxBCG J217.95869+13.53470 in units of keV. (b) Electron number density ($\times 10^{-2}\,\, \rm{cm^{-3}}$). (c)~Quasi-deprojected pressure map ($\times 10^{-11}$ Ba). (d) Entropy map (\keVscm). North is up, east to the left. Radio contours as in~Figure~\ref{fig:clusterimage}.}
 \label{fig:maps}
\end{figure*}

The projected temperature map, with radio contours overlaid, is presented in Figure \ref{fig:maps}. Hot regions are present to the north and west of the BCG. It is interesting to note that the northern hot plasma is parallel to the northern edge of the central radio source, while the western hot plasma is approximately collinear with the south-western radio emission.

\subsection{Pressure and entropy maps}

We used the results of the spectral fits to create quasi-deprojected pressure and entropy maps of the cluster. 

We calculated the electron number density corresponding to each bin using the definition of the spectrum normalization,
\begin{eqnarray}
  \mathcal{N}_{\rm bin} & = & \frac{10^{-14}}{4\pi \left[D_{\rm A}(1+z)\right]^2}\int n_{\rm e} n_{\rm H} {\rm d}V \nonumber \\
              & \approx & \frac{10^{-14}}{4\pi \left[D_{\rm A}(1+z)\right]^2} n_{\rm e} n_{\rm H} V_{\rm bin} \,,
\end{eqnarray}
where $D_{\rm A}$ is the angular size distance, $D_{\rm A}=568.9$ Mpc, $n_{\rm H}$ is the hydrogen number density, $V_{\rm bin}$ is the bin volume, and we assume $n_{\rm e}/n_{\rm H} \approx 1.2$.

Since the normalization parameters were left to vary independently in our previous fits, we refitted the spectra with a Galactically-absorbed single-temperature \texttt{APEC} model in which the temperature and the metallicity were fixed to the previously determined values, while the normalizations were left free to fit, but were constrained to be the same for all detectors.

To estimate the volume of each bin, we assume that the observed emission originates only from the region between the spheres of minimum and maximum radii tangent to the bin \citep{Mahdavi_etal_2005}. Let $R_{\rm max}$ and $R_{\rm min}$ be the maximum and minimum radii, respectively. The volume of a bin is then
\begin{equation}
  V = \frac{2SL}{3} \,,
\end{equation}
where $S$ is the projected area of the bin in the sky plane, and $L=2\sqrt{R_{\rm max}^2-R_{\rm min}^2}$ is the longest length through the contributing volume. The projected area can be easily calculated by directly counting the number of pixels in the bin, and using the fact that the pixel scale is 4 arcsec/pixel. With this method of describing the 3D geometry of a bin, we are considering a constant temperature along the line of sight. 

The electron number density map is shown in Figure \ref{fig:maps}. At the 1-sigma level, the uncertainties are 3-5\%.

Using the bin electron number density and temperature, we calculated the pressure, $p=n_{\rm e}kT_X$, and entropy, $K=kT_{\rm X} n_{\rm e}^{-2/3}$, where $k$ is Boltzmann's constant, and $T_X$ is the temperature of the X-ray emitting gas. The two maps are shown in Figure \ref{fig:maps}. The uncertainties at the 1-sigma level are 9-20\% and 8-20\%, respectively, for pressures and entropies. 

A pressure gradient is seen in the pressure map, with the pressure decreasing uniformly toward the outskirts. The entropy distribution follows a less clear pattern. The south-western radio source coincides with a region of higher entropy. High entropy regions are also present west, northeast and northwest of the BCG.

\begin{figure*}
 \center
 \includegraphics[scale=0.47]{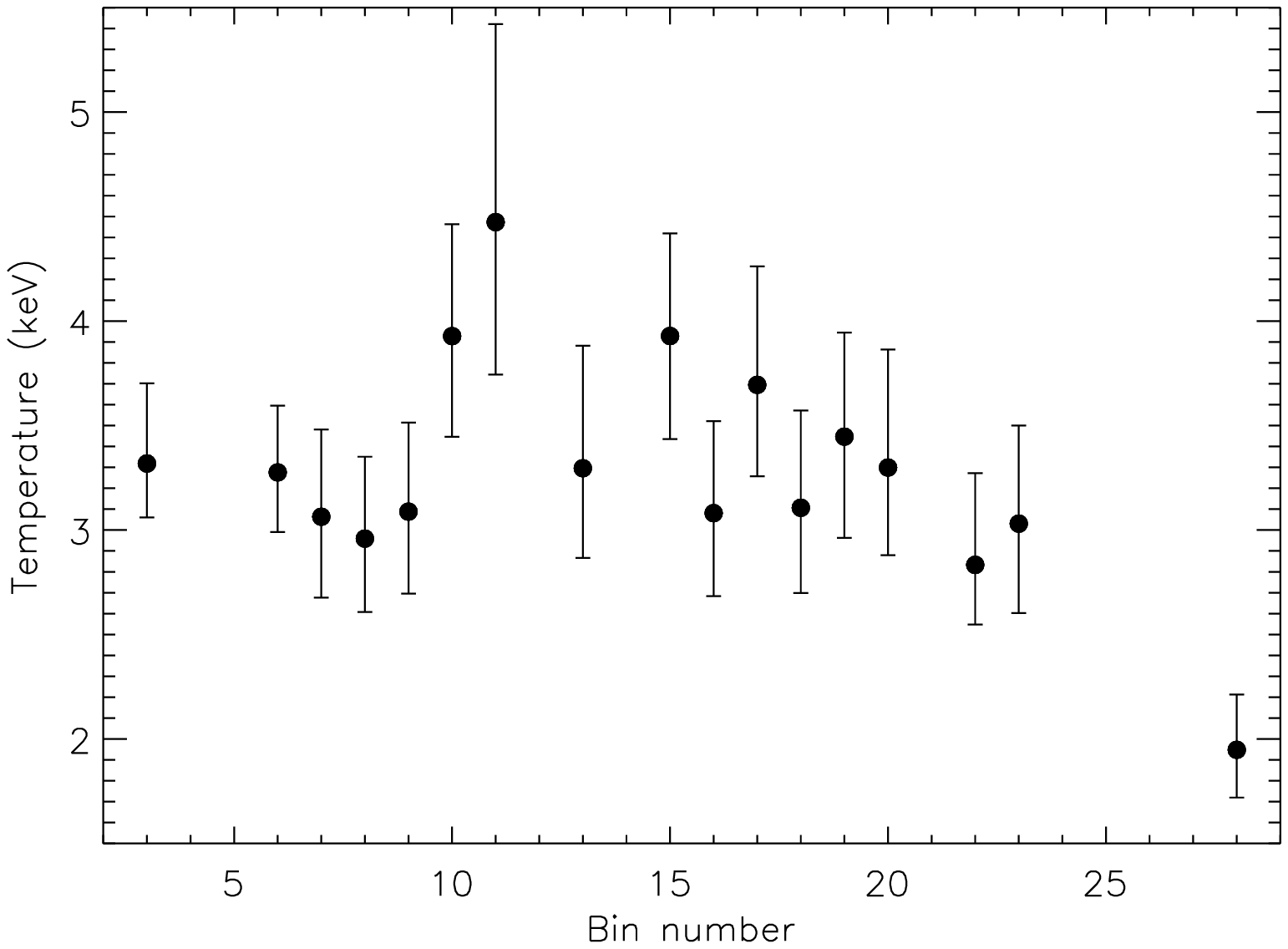}
 \includegraphics[scale=0.47]{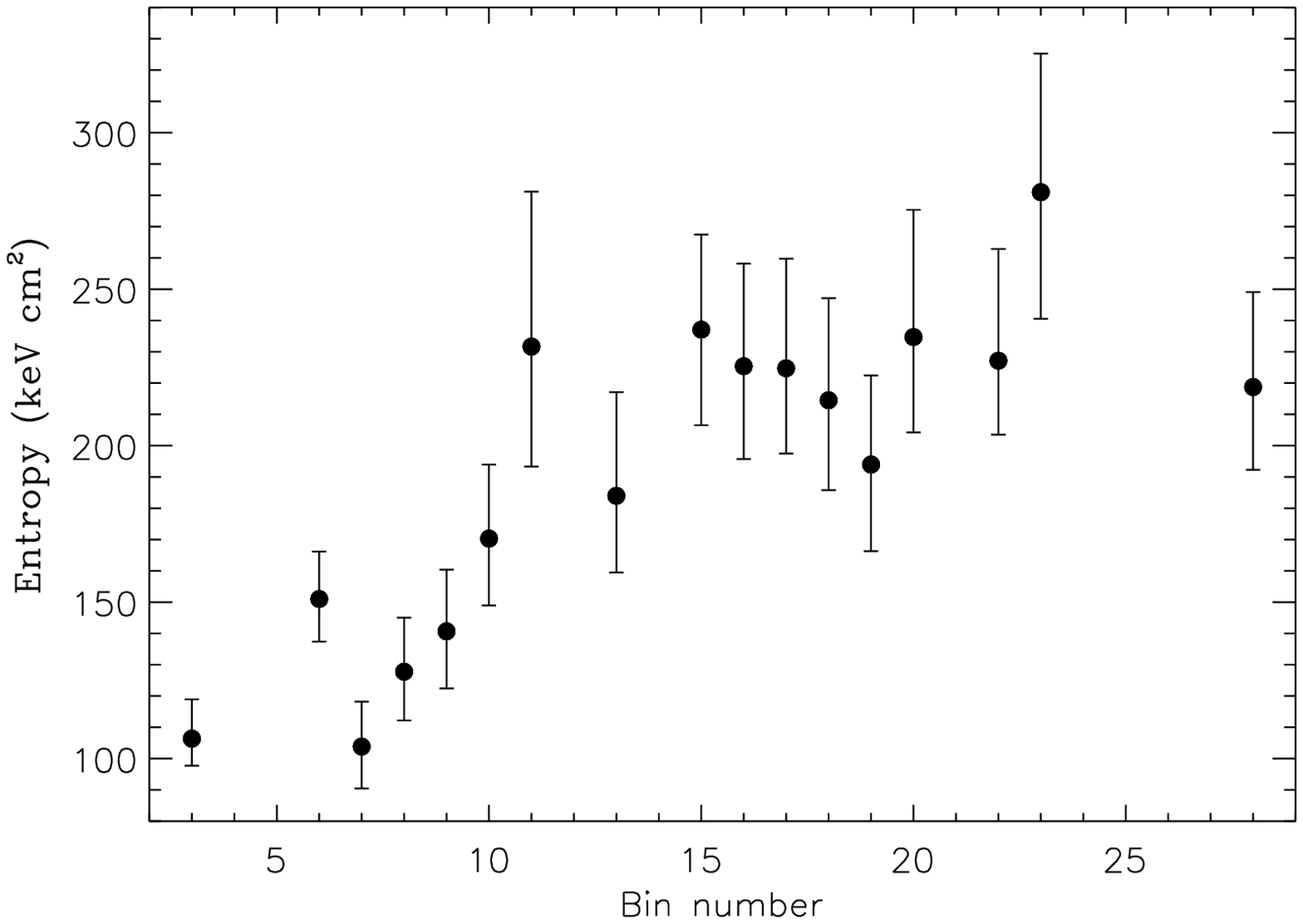}
 \caption{\emph{Left:} Temperatures in the bins shown in Figure \ref{fig:binno}. \emph{Right:} Entropies in the bins shown in Figure \ref{fig:binno}. Error bars are plotted at the 1$\sigma$ level.}
 \label{fig:compare}
\end{figure*}

\begin{figure}
 \center
 \includegraphics[scale=0.016]{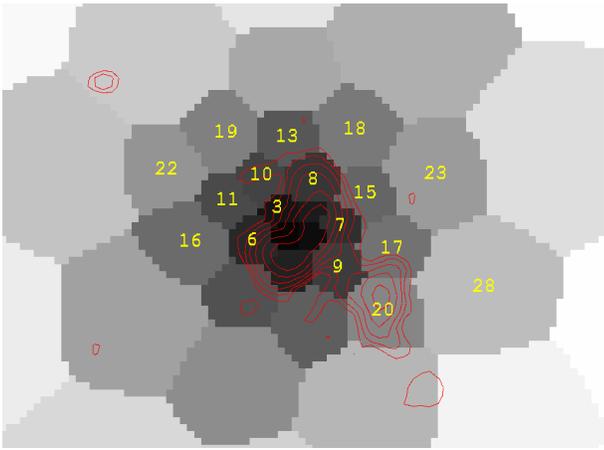}
 \caption{Bin numbers of some of the central regions in the cluster. Temperatures and entropies for these bins are shown in Figure~\ref{fig:compare}. Radio contours as in~Figure \ref{fig:clusterimage}.}
 \label{fig:binno}
\end{figure}

Figure \ref{fig:compare} shows the temperatures and entropies, plotted against bin number, in most of the bins covering the center of the cluster. The bin numbers are shown in Figure \ref{fig:binno}. The bins were selected based on the variations observed in the temperature and entropy maps.

\section{Results}
\label{sec:results}

\subsection{Mass}

\begin{figure}
 \center
 \includegraphics[width=\columnwidth]{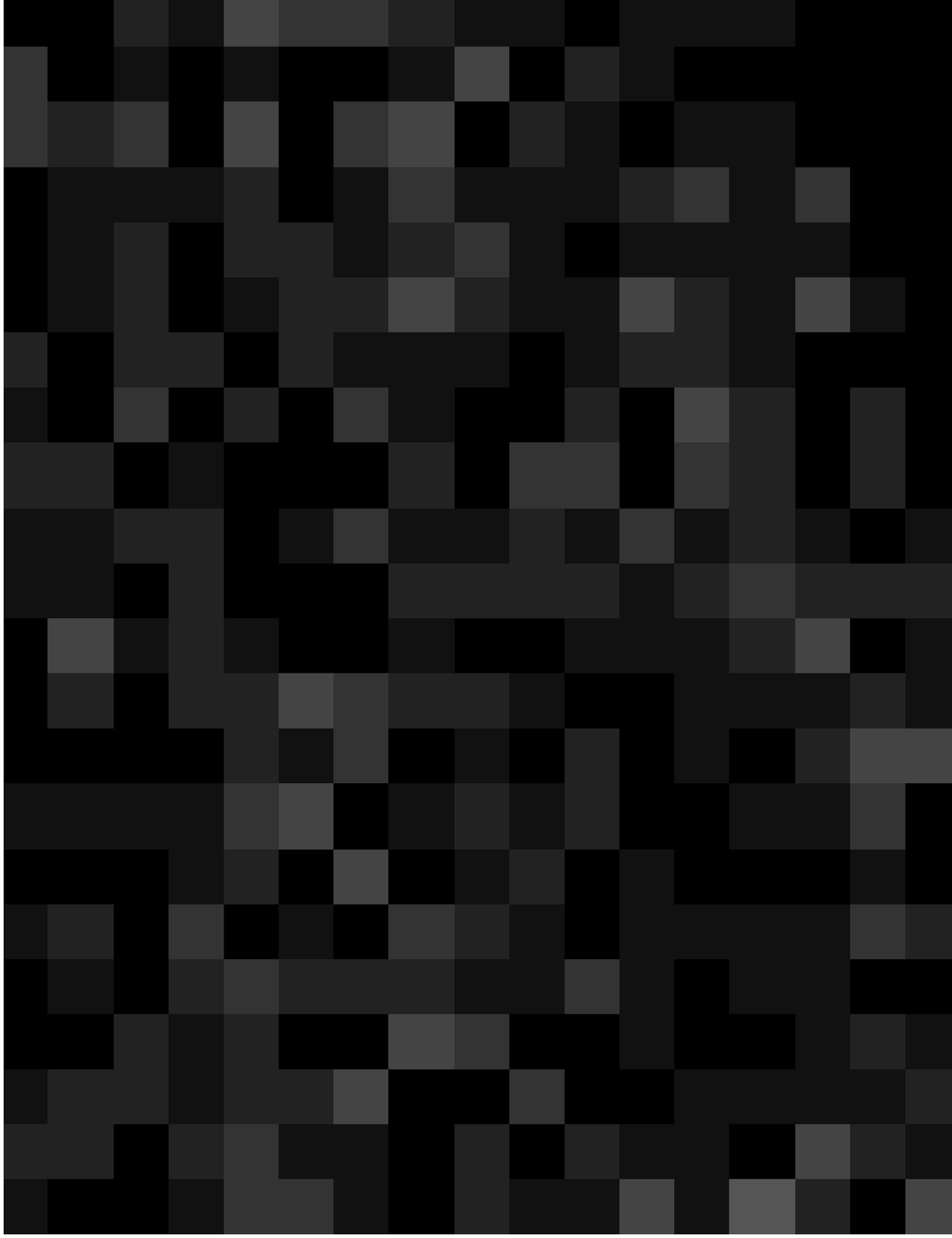}
 \caption{Concentric sector annuli used for creating temperature and electron number density profiles. The sector was chosen such that it does not cross the hot regions visible in the temperature map of the cluster. The position of the AGN that was not included in the analysis is marked by a green circle; the extracted region has a radius of 20 arcsec.}
 \label{fig:annuli}
\end{figure}

To calculate the mass profile of the cluster, we extracted spectra from concentric sector annuli centered on the BCG (Fig. \ref{fig:annuli}). This sector was chosen to be separated from the hot regions found in the temperature map. As before, the spectra were grouped to a minimum of 30 counts per spectral bin, and fitted with an absorbed single-temperature APEC model. The results of these fits are summarized in Table \ref{tab:annuli}. The normalization of the model, together with the volume matrix \citep{Kriss1983} allow us to create the deprojected electron number density profile of the cluster using the geometrical deprojection model of \cite{Ettori2002}.

\begin{center} 
 \begin{table}
  \caption{Radii, temperature, normalization of the best-fit APEC model, and reduced $\chi^2$ of the APEC fit for the sector annuli regions shown in Figure \ref{fig:annuli}. Errors quoted at the 1$\sigma$ level.}
 \begin{tabular}{|c|c|c|c|} \hline 
   Radius (arcsec) & $T$ (keV) & $\mathcal{N}\times 10^5$ & $\chi^2$/d.o.f. \\ \hline\hline
   $0-20$ & $3.2_{-0.48}^{+0.60}$ & $5.2\pm 0.22$ & $0.66$ \\ \hline 
   $20-35$ & $3.3_{-0.55}^{+0.79}$ & $4.8\pm 0.23$ & $0.52$ \\ \hline 
   $35-55$ & $2.6_{-0.30}^{+0.39}$ & $8.3\pm 0.31$ & $0.49$ \\ \hline 
   $55-80$ & $3.1_{-0.49}^{+0.51}$ & $7.5\pm 0.30$ & $0.48$ \\ \hline 
   $80-110$ & $2.5_{-0.44}^{+0.48}$ & $6.7\pm 0.30$ & $0.64$ \\ \hline 
   $110-160$ & $1.3_{-0.10}^{+0.10}$ & $12\pm 0.55$ & $0.80$ \\ \hline 
   $160-210$ & $0.98_{-0.070}^{+0.069}$ & $8.7\pm 0.53$ & $0.69$ \\ \hline 
   $210-285$ & $0.96_{-0.061}^{+0.054}$ & $6.6\pm 0.42$ & $0.87$ \\ \hline 
  \label{tab:annuli}
 \end{tabular}
 \end{table} 
\end{center}

\begin{figure}
 \center
 \includegraphics[width=\columnwidth]{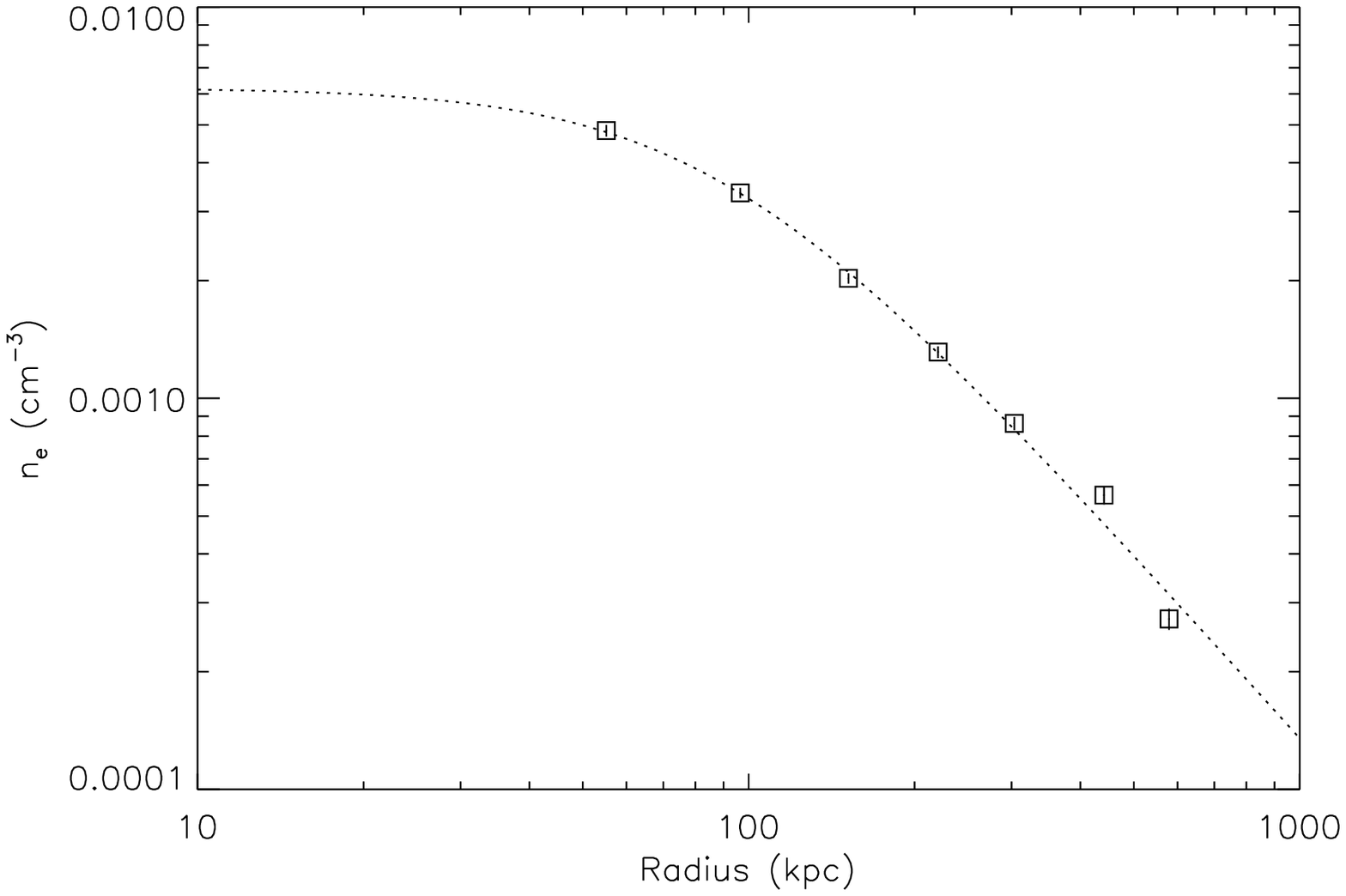}
 \includegraphics[width=\columnwidth]{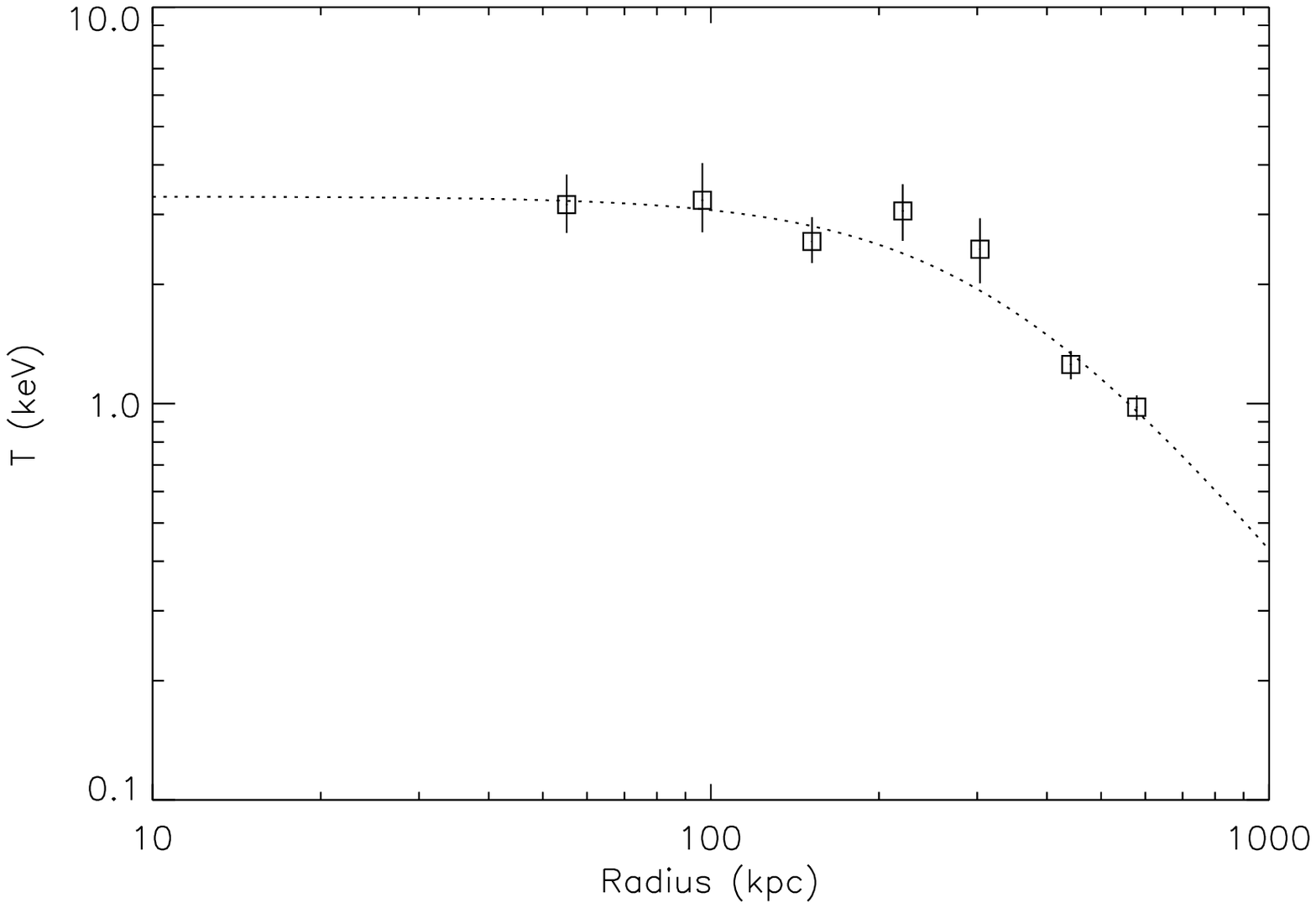}
 \caption{Deprojected electron number density (top) and projected temperature (bottom) profiles of the cluster, extracted from the sector shown in Fig. \ref{fig:annuli}. The models fitted to these profiles are shown in red. Error bars are plotted at the 1$\sigma$ level.}
 \label{fig:profiles}
\end{figure}

The deprojected electron number density profile was fitted with a standard beta-model, yielding the best-fit parameters $\beta=0.5\pm 0.02$, $r_c=88\pm 8$ kpc, $n_0=0.006\pm 0.0004$ \pccm, while the projected temperature profile was fitted with a model of the form:
\begin{eqnarray}
  T = \frac{T_0}{\left[1+\left(\frac{r}{r_\delta}\right)^2\right]^{\delta}}.
\end{eqnarray}
The best-fit parameters for the temperature model are $T_0=3.3\pm 0.4$ keV, $r_\delta=326\pm 196$ kpc, $\delta=0.9\pm 0.5$. The profiles and the fitted models are shown in Fig. \ref{fig:profiles}.

We first determined the mass of the cluster by integrating over the electron number densities in the sector in Fig. \ref{fig:annuli}, up to $R_{500} = 527$ kpc (corresponding to the radius where the average density is 500 times the critical density of the universe). The gravitational mass is given by $M = M_{\rm gas}/f_b$, with the mass of the gas $M_{\rm gas} = \int \mu_e m_{\rm H}n_e dV$, where $\mu_e=1.17$ is the mean molecular weight per electron, and the baryon fraction $f_b$. Using the gas mass fraction inside $R_{500}$ from \cite{Vikhlinin2006}, and correcting for stellar components \citep{Gonzalez2007}, the gravitational mass of the cluster within $R_{500}$ is $9.2\times 10^{13}$ \Msun. If the baryon fraction of the cluster approaches the mean cosmic baryon fraction \citep[0.167; ][]{Komatsu2010}, then this sets a lower limit of $5.8\times 10^{13}$ \Msun\, on the gravitational mass.

\begin{figure}
 \center
 \includegraphics[width=\columnwidth]{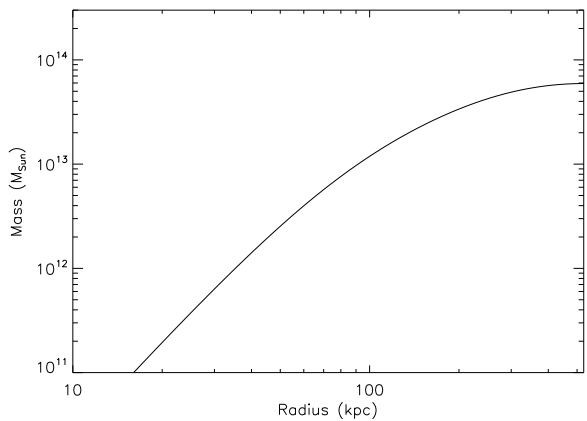}
 \caption{Gravitational mass profile of the cluster within $R_{500}$, derived from the fits to the temperature and electron number density profiles.}
 \label{fig:massprofile}
\end{figure}

By assuming hydrostatic equilibrium and spherical symmetry, the gravitational mass of the cluster can also be calculated from the models fitted to the electron number density and temperature profiles via:
\begin{eqnarray}
  M(r) = -\frac{kTr}{G\mu m_{\rm H}} \left[\frac{d \log n_e}{d \log r}+\frac{d \log T}{d \log r}\right]\,,
\end{eqnarray}
where $T$ is the X-ray temperature of the gas at radius $r$, $k$ is Boltzmann's constant, $G$ is the gravitational constant, $\mu\approx 0.6$ is the mean molecular weight of the ionized plasma, and $m_{\rm H}$ is the hydrogen mass. The resulting mass profile is shown in Fig. \ref{fig:massprofile}. The mass enclosed in a radius of $R_{500}$ is $5.9\times 10^{13}$ \Msun.

Alternatively, the mass can be estimated from the $M_{500}-Y_{\rm X}$ scaling relation, where $Y_{\rm X} = M_{\rm gas,500}T_{500}$, where $T_{500}$ is the X-ray temperature of the gas in the $[0.15-1]R_{500}$ region \citep{Pratt2009} of the sector showed in Fig. \ref{fig:annuli}. 

Cluster spectra were extracted for all detectors from radius $0.15R_{500}$ up to $R_{500}$ in the sector in Fig. \ref{fig:annuli}, and rebinned to a minimum of 30 counts per spectral bin, to allow for $\chi^2$ statistics. From the fits to the spectra, which were done following the same method as for the bin spectra, we find a temperature $T_{500}=2.1$ keV, while the deprojected electron number density profile yields a gas mass $M_{\rm gas, 500} = 1.2\times 10^{13}$ \Msun within $R_500$. Therefore, using equation (1) in \cite{Pratt2009}, the gravitational mass of the cluster within $R_{500}$ is $\sim 1.1\times 10^{14}$ \Msun. 

Spectra were also extracted from the full sector in Fig. \ref{fig:annuli} up to $R_{500}$ (i.e., from radius 0 to $R_{500}$), and fitted with an absorbed \texttt{APEC} model in the same way as described in Section \ref{ssec:tmap}. The fit yields a flux of $\sim 1.1\times 10^{-12}$ \ergpscmps in the $0.5-2$ keV energy range, equivalent to a luminosity of the cluster of $7.9\times 10^{43}$ \ergps. Using the $L[0.5-2]-M_{500}$ relation from \cite{Pratt2009}, this luminosity corresponds to a total mass within $R_{500}$ of $1.8\times 10^{14}$ \Msun.

The disturbed temperature distribution suggests that \j1431 is possibly a system undergoing a merger. By selecting the extraction region of our spectra such as not to include the hot northern and western regions, our aim was to minimize the influence of a possible merger on the different mass estimates. However, the assumptions of hydrostatic equilibrium and spherical symmetry do not necessarily hold. The mass calculated under the assumption of hydrostatic equilibrium could be underestimated due to residual non-thermal support \citep[see e.g., ][]{Zhang2008}. The mass inferred from the $M_{500}-Y_{\rm X}$ and $L[0.5-2]-M_{500}$ scaling relations could presumably also be affected by the interaction of \j1431 with another galaxy cluster. Nevertheless, we believe the mass of the cluster to be closer to $1\times 10^{14}$ \Msun, higher than the mass calculated under the assumption of hydrostatic equlibrium.

\subsection{Central cooling time}

The cooling time of optically thin gas is given by
\begin{equation}
t_{\rm cool} = \frac{3nkT_{\rm X}}{(\gamma-1)n_{\rm e} n_{\rm H} \Lambda (T_{\rm X},Z)}\,,
\end{equation}
where $n$ is the total number density ($n\approx 2.3 n_{\rm H}$), $\gamma$ is the adiabatic index ($\gamma=5/3$), and $\Lambda (T_{\rm X},Z)$ is the cooling function for temperature $T_{\rm X}$ and metallicity $Z$. Adopting the cooling function from \cite{SutherlandDopita1993}, the central cooling time of the cluster is found to be $6.1 \pm 0.53$ Gyr. This indicates that \j1431 is a weak cool core or transition cluster \citep{Hudson2010}.

However, owing to the point-spread functions (PSFs) of the \emph{XMM-Newton} telescopes, the central X-ray luminosity may be more peaked, which would lower the central cooling time. Therefore, $6.1$ Gyr is only an upper limit. Observations with the \emph{Chandra} telescope would be necessary to see if \j1431 is a stronger cool core.

\section{Discussion}
\label{sec:discussion}

\subsection{The dynamical state of the cluster}

\begin{figure}
 \center
 \includegraphics[width=\columnwidth]{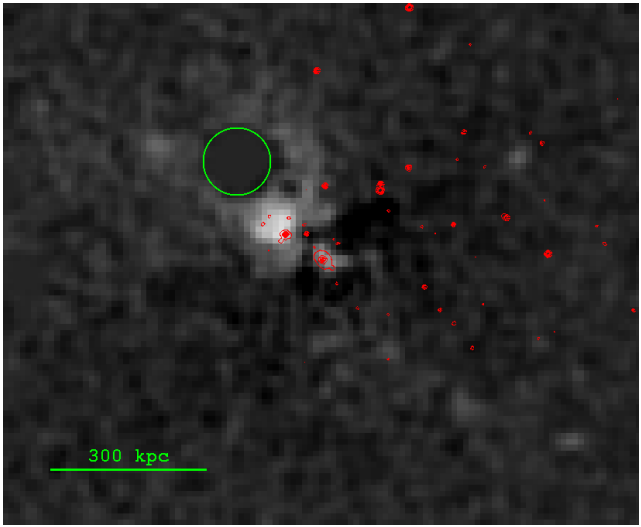}
 \caption{After subtraction of a symmetric brightness profile, excess X-ray emission can be easily observed to the NE of the cluster center. Overlaid in red are SDSS optical contours. The northeastern AGN was excluded from the image, and the excluded region is marked by a green circle of radius 20 arcsec.}
 \label{fig:structure}
\end{figure}

To search for evidence of X-ray excess in the central region of the cluster, we calculated an azimuthally-averaged surface brightness profile by considering 50 $8^{\prime\prime}$-wide rings centered on the brightest X-ray pixel. This profile was used to construct a symmetric cluster image, which was subsequently subtracted from the original image of the cluster. The result of this subtraction is shown in Fig. \ref{fig:structure}. After subtraction, an X-ray excess can be seen $\sim 90$ kpc to the NE of the cluster center. This excess is associated with a few galaxies visible in the SDSS optical image, and suggests an interaction of \j1431 with another galaxy cluster or group of galaxies.

This hypothesis is further supported by the disturbed temperature distribution, and by the bent radio emission, all of which point toward a disturbed, merging system. 

To estimate the mass of the second galaxy cluster, an azimuthally-averaged surface brightness profile of \j1431 was created while masking the emission from the second cluster. Subtracting this profile from the original image allows us to put a lower limit on the luminosity of the second cluster. The ratio of the luminosities of the two clusters is $\sim 6$. Assuming the clusters are in the same dynamical state and using the $L_{500}-M_{500}$ scaling relations from \cite{Pratt2009}, the mass of the second cluster is $10-40\%$ of the mass of \j1431, with the lower value corresponding to the scaling relation of a disturbed system.

Assuming the merging process began at z=0.2 and the two clusters dominate the mass in their region of the Universe, the greatest separation between them is $d_0 \sim 2.5$ Mpc. The exact age of the Universe at the time of the merger has very little effect on the relative velocity and on the impact parameter. Using equations (23) and (19) in \cite{Sarazin2002}, the impact parameter of the smaller cluster can be estimated via
\begin{equation}
b \approx \lambda \sqrt{\frac{d_0d}{2}}\left(1-\frac{d}{d_0}\right)^{-1/2} f\left(M_1,M_2\right)\,,
\end{equation}
where $\lambda$ is the spin parameter \citep{Peebles1969}, $d$ is the current separation between clusters, and
\begin{equation}
f\left(M_1,M_2\right) = \frac{(M_1+M_2)^3}{M_1^{3/2}M_2^{3/2}} \left[1-\frac{\left(M_1^{5/3}+M_2^{5/3}\right)}{(M_1+M_2)^{5/3}}\right]^{3/2}\,,
\end{equation}
where $M_1$ and $M_2$ are the masses of the two clusters. 

The relative velocity of the clusters can be estimated from equation (14) of \cite{Sarazin2002}:
\begin{equation}
v^2 \approx 2G\,(M_1+M_2)\left(\frac{1}{d}-\frac{1}{d_0}\right)\,\left[1-\left(\frac{b}{d_0}\right)^2\right]^{-1}\,.
\end{equation}

We assume a spin parameter $\lambda = 0.05$ \citep{Sarazin2002}. If the mass of \j1431 is $\sim 1\times 10^{14}$ \Msun\, and if the current separation between the clusters' centres is equal to the projected distance between them, $d \sim 90$ kpc, the impact parameter for a disturbed system with a mass ratio of 1:10 is $\sim 30$ kpc, while the relative velocity of the clusters is $\sim 3200$ \kmps. This velocity is approximately six times higher than the sound velocity within $R_{500}$ ($c_s=568$ \kmps), and this is possibly only a lower limit since the luminosity (hence, the mass) of the smaller cluster is likely to be underestimated. It is possible, however, to overestimate the relative velocity due to projection effects. If we assume that the temperature increase observed in the north equals the temperature jump across a merger shock, then the shock has a Mach number of 1.3, which corresponds to a shock velocity of $\sim 1000$ \kmps. For this velocity to be comparable to the relative velocity of the merging clusters, the separation between them should be $\sim 700$ kpc. This would yield an impact parameter of $\sim 100$ kpc. 

We speculate that the smaller cluster entered the ICM of \j1431 from southeast, giving rise to the high-temperature structures visible north and west from the BCG. If this is true, and the separation between the BCGs of the two clusters is small, then we are likely observing the system shortly after closest encounter.

\subsection{Radio relic}

The southwestern radio source is not associated with X-ray or optical emission. However, there is a bridge of faint radio emission connecting it to the central radio source, suggesting that it is related to the radio AGN. The radio emission of the southwestern source is elongated and bent eastward, which implies that the relic is being acted upon by the surrounding ICM. Its very curved radio spectrum, with $\alpha=-1.5$ between 325 and 610 MHz, steepening to $\alpha=-2.5$ between 610 and 1425 MHz, indicates that the plasma has undergone significant synchroton and inverse Compton loses. Unless the source has been ``revived'' by the passing of a shock capable of compressing the faded plasma, the southwestern source would probably not be visible even at the relatively low GMRT frequencies. \cite{Ensslin2001} and \cite{Ensslin2002} have shown that the energy gained during adiabatic compression, combined with the increase in magnetic field strength, can cause the aged plasma to emit radio waves again. Given the signs of merging present in the X-ray observations, we speculate that the southwestern radio source originated as a bubble from the central AGN, lost significant energy via synchroton and inverse Compton emission, and was subsequently revived by a merger shock triggered by the interaction between \j1431 and a smaller cluster or group of galaxies. Therefore, we classify the source as a radio phoenix.

The simplest velocity estimate of a rising bubble can be obtained by equating the ram pressure and buoyancy forces acting on a bubble \citep[see, e.g., ][]{Gull1973}. This yields a terminal velocity of the bubble of 
\begin{equation}
v\sim \sqrt{g\frac{4r}{3}\frac{2}{C}} ,
\end{equation}
where $r$ is the bubble radius, $g$ the gravitational acceleration, and $C$ the drag coefficient.  The expression for the terminal
velocity can be further rewritten using the Keplerian velocity at a given distance from the cluster center: $v\sim[(r/R)(8/3C)]^{1/2}v_K$, where $R$ is the distance from the center, and $v_K = \sqrt{gR}$ is the Keplerian velocity. For a solid sphere moving through an incompressible fluid, the drag coefficient $C$ is of the order of 0.4-0.5 for Reynolds numbers in the range of $10^3-10^5$ \citep[see, e.g., ]{Landau1959}. Thus, a large and strongly underdense bubble will rise with a velocity comparable to the Keplerian velocity.

In MaxBCG J219.95869+13.53470 the radio source in the SW is at a distance of $\sim 1$ arcmin from the central AGN which corresponds to
a distance of 165 kpc. The mass enclosed within this radius, calculated by integrating over density, is $1.2\times 10^{13}$ \Msun, so so the Keplerian velocity 165 kpc the BCG is $v_K \sim 550$ \kmps. Therefore, it would take at least $300$ Myr for the bubble to rise to its current location. However, the synchrotron lifetime of the electrons is less than approximately 100 Myr -- shorter than the buoyant rise time.

In cluster centres, Mach numbers of merger shocks are low and hence compression factors inside shock fronts are unlikely to exceed $2-3$. The compression caused by the merger shock wave gives rise to a burst of low frequency emission, but practically no high frequency emission. This is due to the rapid decay of the upper end of the electron spectrum in the dense cluster centres, which essentially wipes out the adiabatic energy gains of these electrons. The source decays on a time-scale of a few tens of Myr, mostly due to the heavy synchrotron losses.

The passage of a merger shock wave through the intracluster medium causes irreversible transformations that lead to an increase in entropy. In \j1431, the entropy map shows an overall gradual increase in entropy from the centre toward outskirts. However, on top of this natural entropy gradient, there are several areas of higher entropy surrounding the central radio source. One of these areas coincides with the position of the southwestern radio source, and it is possibly a result of the processes leading to the revival of old plasma.

\section{Conclusions}
\label{sec:conclusions}

SDSS optical images of \j1431 reveal the presence of two bright galaxies at the center of the cluster. The BCG of the galaxy cluster is located at a redshift of 0.16, while the second galaxy is at a redshift of 0.162.~\footnote{http://www.sdss.org/DR7/} Both galaxies are visible in X-ray, at a projected distance of $\sim 90$ kpc. Based on the ratio of their X-ray luminosities, the smaller galaxy appears to be the BCG of a second galaxy cluster with a mass of $10-40$ percent of the mass of \j1431, and given the similar redshifts of the two galaxies, they are likely interacting. Using a simple merger model, we estimate the impact parameter to be $\sim 30-100$ kpc.

Using \emph{XMM-Newton} observations of \j1431, we characterize the physical properties and the dynamical state of the cluster. A single-temperature \texttt{APEC} spectral fit to the $R_{500}$ region yields a luminosity of $\sim 7.9\times 10^{43}$ \ergps in the $0.5-2$ keV energy range. We also constructed temperature, electron number density, pressure and entropy maps of the cluster. To the north and west of the BCG, the temperature map shows the presence of hot plasma, with temperatures $\sim 1.5-2$ times higher than the temperature of its surroundings. The northern hot region lies right to the northwest of the smaller galaxy. 

Based on the temperature map, we created projected temperature and deprojected electron number density profiles in a sector away from the hot region, and we fitted these profiles with a temperature model and a beta-model, respectively. From the fit to the deprojected electron number density profile we estimate the mass of the cluster assuming a baryon fraction $f_b = 0.133$ \citep{Gonzalez2007}, and find a value of $9.2\times 10^{13}$ \Msun. Assuming spherical symmetry and hydrostatic equilibrium, the mass estimated from the beta-model and the temperature model is $5.9\times 10^{13}$ \Msun. Excluding the central $0.15R_{500}$ region, we use the scaling relation between $Y_{\rm X}$ and $M_{500}$ to find $M_{500} \sim 1.1\times 10^{14}$ \Msun. From the $L[0.5-2]-M_{500}$ scaling relation of \cite{Pratt2009}, the mass within $R_{500}$ is $1.8\times 10^{14}$ \Msun. Considering that we are observing a disturbed system, we believe that the mass is higher than calculated from hydrostatic equilibrium, so it is probably closer to $1\times 10^{14}$ \Msun.

The regions of high temperature seen in the temperature map are also characterized by high entropies. Furthermore, to the west of the western region, there is a region of anomalously high entropy compared to its surroundings. Another region of high entropy can be seen to the south-west of the BCG. The hot regions, as well as the disturbed entropy distribution, support the hypothesis of a merger.

The southwestern high entropy region coincides with the position of the southwestern radio emission visible in the GMRT images. The southwestern radio source is connected to the central radio source by a bridge of faint radio emission. Therefore, it is likely that the two sources are connected, and the southwestern source traces its origin to the center of the cluster. Based on the radio data of \cite{vanWeeren2009} and on the evidence of a merger from the X-ray observations, we speculate that the southwestern radio source originated as a bubble from the central AGN, lost significant energy via synchroton and inverse Compton emission, and was subsequently revived by the merger of \j1431 with the second galaxy cluster. This smaller cluster appears to have entered the ICM of \j1431 from southeast, the two clusters being currently observed just after the closest encounter of their BCGs.

Higher resolution X-ray observations and integral field spectroscopy data of \j1431 are needed to provide more information about the geometry and state of the merger.

\section*{Acknowledgments}

This work is based on observations obtained with XMM-Newton, an ESA science mission with instruments and contributions directly funded by ESA Member States and the USA (NASA). MB acknowledges support by the research group FOR 1254 funded by the Deutsche Forschungsgemeinschaft. RJvW acknowledges funding from the Royal Netherlands Academy of Arts and Sciences. AS was supported by Einstein Postdoctoral Fellowship grant number PF9-00070 awarded by the Chandra X-ray Center, which is operated by the Smithsonian Astrophysical Observatory for NASA under contract NAS8-03060. JHC acknowledges the support of the South-East Physics Network (SEPNet).

\bibliographystyle{mn2e}
\bibliography{bibliography}

\label{lastpage}

\end{document}